\documentclass[prodmode,acmtecs]{acmsmall} % Aptara syntax

\usepackage[ruled]{algorithm2e}

\usepackage{epstopdf}
\usepackage{graphics} % for EPS, load graphicx instead
\usepackage{times}    % comment if you want LaTeX's default font
\usepackage{url}      % llt: nicely formatted URLs
\usepackage{comment}
\usepackage{multirow}
\usepackage{caption}
\usepackage{tablefootnote}
\usepackage{color}
\usepackage{amsmath}

\SetAlFnt{\small}
\SetAlCapFnt{\small}
\SetAlCapNameFnt{\small}
\SetAlCapHSkip{0pt}
\IncMargin{-\parindent}

\begin{document}

% Page heads
\markboth{C. Dong et al.}{PPM: A Privacy Prediction Model for Online Social Networks}

% Title portion
\title{PPM: A Privacy Prediction Model for Online Social Networks}
\author{
	CAILING DONG
	\affil{University of Maryland, Baltimore County}
	HONGXIA JIN
	\affil{Samsung Research America}
	BART P. KNIJNENBURG
	\affil{Clemson University}
}
% NOTE! Affiliations placed here should be for the institution where the
%       BULK of the research was done. If the author has gone to a new
%       institution, before publication, the (above) affiliation should NOT be changed.
%       The authors 'current' address may be given in the "Author's addresses:" block (below).
%       So for example, Mr. Abdelzaher, the bulk of the research was done at UIUC, and he is
%       currently affiliated with NASA.

\begin{abstract}
Online Social Networks (OSNs) have come to play an increasingly important role in our social lives, and their inherent privacy problems have become a major concern for users. Can we assist consumers in their privacy decision-making practices, for example by predicting their preferences and giving them personalized advice? To this end, we introduce PPM: a Privacy Prediction Model, rooted in psychological principles, which can be used to give users personalized advice regarding their privacy decision-making practices. Using this model, we study psychological variables that are known to affect users' disclosure behavior: the \emph{trustworthiness} of the requester/information audience, the \emph{sharing tendency} of the receiver/information holder, the \emph{sensitivity} of the requested/shared information, the \emph{appropriateness} of the request/sharing activities, as well as several more traditional \emph{contextual factors}.
\end{abstract}

\begin{CCSXML}
<ccs2012>
<concept>
<concept_id>10002978.10003022.10003027</concept_id>
<concept_desc>Security and privacy~Social network security and privacy</concept_desc>
<concept_significance>300</concept_significance>
</concept>
<concept>
<concept_id>10002978.10003029.10011150</concept_id>
<concept_desc>Security and privacy~Privacy protections</concept_desc>
<concept_significance>300</concept_significance>
</concept>
<concept>
<concept_id>10003120.10003130.10003134.10003293</concept_id>
<concept_desc>Human-centered computing~Social network analysis</concept_desc>
<concept_significance>100</concept_significance>
</concept>
</ccs2012>
\end{CCSXML}

\ccsdesc[300]{Security and privacy~Social network security and privacy}
\ccsdesc[300]{Security and privacy~Privacy protections}
\ccsdesc[100]{Human-centered computing~Social network analysis}


% We no longer use \terms command
%\terms{Design, Algorithms, Performance}

\keywords{Privacy prediction; decision making; classification; machine learning; Online Social Networks (OSNs)}

%\acmformat{Cailing Dong, Hongxia Jin, Bart P. Knijnenburg, 2015. PPM: A Privacy Prediction Model for Online Social Networks.}
% At a minimum you need to supply the author names, year and a title.
% IMPORTANT:
% Full first names whenever they are known, surname last, followed by a period.
% In the case of two authors, 'and' is placed between them.
% In the case of three or more authors, the serial comma is used, that is, all author names
% except the last one but including the penultimate author's name are followed by a comma,
% and then 'and' is placed before the final author's name.
% If only first and middle initials are known, then each initial
% is followed by a period and they are separated by a space.
% The remaining information (journal title, volume, article number, date, etc.) is 'auto-generated'.

%\begin{bottomstuff}
%This work is supported by the National Science Foundation, under 
%grant CNS-0435060, grant CCR-0325197 and grant EN-CS-0329609.

%Author's addresses: C. Dong, Department of Information Systems, University of Maryland, Baltimore County; 
%H. Jin, Samsung Research America; 
%B. Knijnenburg, Human-Centered Computing, Clemson University.
%\end{bottomstuff}

\maketitle

\section{Introduction}
The rising popularity of Online Social Networks (OSNs) has ushered in a new era of social interaction that increasingly takes place online. Pew Research reports that 72\% of online American adults maintain a social network profile~\cite{onlineArticle2013}, which provides them with a convenient way to communicate online with family, friends, and even total strangers. To facilitate this process, people often share personal details about themselves (e.g. likes, friendships, education and work history). Many users even share their current activity and/or real-time location. However, all this public sharing of personal and sometimes private information may increase security risks (e.g., phishing, stalking~\cite{Gross2005,AAHasib2009}), or lead to threats to one's personal reputation~\cite{Baker2011}. It is therefore no surprise that privacy aspects of OSN use has raised considerable attention from researchers, OSN managers, as well as users themselves.

The privacy dilemma OSN users face is the choice between sharing their information (which may result in social benefits~\cite{ellison2007,Joinson2008,DiMicco2008,Lampe2014}) and keeping it private or restricted to certain users only (thereby protecting their privacy). To help users with this decision, experts recommend giving users comprehensive \emph{control} over what data they wish to share, and providing them with more \emph{transparency} regarding the implications of their decisions~\cite{Acquisti06,Benisch2011,Tang2012,Tang2010,Toch2010}. Advocates of transparency and control argue that it empowers users to regulate their privacy at the desired level: without some minimum level of transparency and control, users cannot influence the risk/benefit tradeoff. Moreover, people can only make an informed tradeoff between benefits and risks if they are given adequate information~\cite{Lederer2004,Sadeh2009}.
 
The privacy decisions on ONSs are so numerous and complex, that users often fail to manage their privacy effectively. Many users avoid the hassle of using the ``labyrinthian'' privacy controls that Facebook provides~\cite{Compano2010,CR2012}, and those who make the effort to change their settings do not even seem to grasp the implications of their own privacy settings~\cite{YabingLiu2011,Madejski2012}. There is strong evidence that transparency and control do not work well in practice, and several prominent privacy scholars have denounced their effectiveness in helping users to make better privacy decisions ~\cite{Barocas2009,Nissenbaum2011,Solove2013}. 

Are there more effective ways to assist consumers in their privacy decision-making practices? A solution that has recently been proposed, is to learn users' privacy preferences and subsequently give them \emph{user-tailored decision support}~\cite{KnijnenburgDecisions2013}. To this end, in this paper we introduce PPM: a comprehensive \emph{privacy prediction model} that can be applied in a multitude of privacy decision making scenarios. The model is \emph{theoretically grounded} in psychological research that has investigated the key variables known to affect users' disclosure behavior. Consequently, PPM can be applied to a wide range of OSNs, and arguably even other privacy-sensitive systems such as e-commerce systems.

A practical application of PPM's predictions would be to provide automatic initial default settings in line with users' privacy preferences (e.g., by default, it discloses Mary's location to her best friends on weekends, but it does not disclose John's location to his boss when he is on vacation). These ``smart defaults'' could alleviate the burden of making numerous complex privacy decisions, while at the same time respecting users' inherent privacy preferences. Balebako et al.~\cite{balebako2011} argue that the help provided via such machine learning systems can be seen as ``adaptive nudges''. Indeed, Smith et al.~\cite{Smith2013} argue that ``Smart defaults can become even smarter by adapting to information provided by the consumer as part of the decision-making process.'' (p. 167)

In this paper we will formally define PPM and then use it to comprehensively study the important psychological and contextual factors that affect privacy decision making on OSNs, with the ultimate goal of assisting users to make appropriate privacy decisions. Specifically, we validate PPM in \emph{multiple scenarios}, testing the effect of several psychological antecedents of information disclosure behavior---the \emph{trustworthiness} of the requester/audience, the \emph{sharing tendency} of the user, the \emph{sensitivity} of the information, the \emph{appropriateness} of the request/disclosure---as well as several more traditional \emph{contextual factors} on data collected on Twitter, Google+, and a location sharing preference study. We provide a comparative evaluation of the importance of each of these factors in determining users' privacy decisions.

\section{Related Work}
\subsection{Privacy Decision Making}
A majority of OSN users takes a pragmatic stance on information disclosure~\cite{Sheehan2002,Taylor2003,Consolvo2005}. These ``pragmatists''~\cite{Westin1998} have balanced privacy attitudes: they ask what benefits they get, and balance these benefits against risks to their privacy interests~\cite{Consolvo2005}. This decision process of trading off the anticipated benefits with the risks of disclosure has been dubbed \emph{privacy calculus}~\cite{Culnan1993,Laufer1977}). In making this tradeoff, these users typically decide to share a subset of their personal information with a subset of their contacts~\cite{Olson2005,Consolvo2005,LewisKevin2008,Lampe2008}.

The term \emph{privacy calculus} makes it sound like users make ``calculated'' decisions to share or withhold their personal information. In reality though, these decisions are numerous and complex, and often involve uncertain or unknown outcomes~\cite{Knijnenburg2013}. Acquisti and Grossklags~\cite{Acquisti2005} identified \emph{incomplete information}, \emph{bounded rationality}, and \emph{systematic psychological deviations from rationality} as three main challenges in privacy decision making. Consequently, people's privacy behavior is far from calculated or rational. Most OSN users share much more freely than expected based on their attitudes~\cite{Acquisti06,Becker2012,BartJinLocation2013,Acquisti2005} (a disparity that has been labeled the ``privacy paradox''~\cite{Norberg2007}). In~\cite{YabingLiu2011}, the authors quantified this disparity between the desired and actual privacy settings, and found that users' privacy settings match users' expectations only 37\% of the time and oftentimes people shared more than they expected.

\subsection{Predicting Privacy Decisions}
When left to their own devices, users thus seem particularly inept at making even the simplest privacy decisions in a rational manner~\cite{BartJinLocation2013}, and many users actively try to avoid the hassle of making such decisions~\cite{Compano2010}. Knijnenburg et al. have recently proposed a way to circumvent users' unwillingness or inability to make accurate privacy decisions: if one can \emph{predict} users' privacy preferences, one can give them \emph{user-tailored decision support}~\cite{KnijnenburgDecisions2013} in the form of recommendations~\cite{KnijnenburgSigHCI2013} or adaptive defaults~\cite{Knijnenburg2014}, which would help to limit the number and complexity of privacy decisions that OSN users have to make. Regarding the first step of this proposal (i.e., predicting users' privacy decisions), scientists have had modest success using various machine learning practices.

For example, Ravichandran et al.~\cite{Ravichandran2009} applied k-means clustering to users' contextualized location sharing decisions to come up with a number of default policies. They showed that a small number of default policies learned from users' contextual location sharing decisions could accurately capture a large part of their location sharing preferences. Similarly, Sadeh et al.~\cite{Sadeh2009} applied a k-nearest neighbor (kNN) algorithm and a random forest algorithm to accurately predict users' privacy preferences in a location-sharing system based on the type of recipient and the time and location of the request. They showed that users had difficulties setting their privacy preferences, and that the applied machine learning techniques could help users in specifying more accurate disclosure preferences. Pallapa et al.~\cite{Pallapa2014} proposed a system that determines the level of privacy required in new situations based on the history of interaction between users. They demonstrated that this solution can deal with the rise of privacy concerns while at the same time efficiently supporting users in a pervasive system full of dynamic and rich interactions. Finally, in a social network context, Fang and LeFevre~\cite{Fang2010} developed a privacy wizard that is able to configure users' privacy settings automatically and accurately with a machine learning model that they developed. The wizard removes the burden of setting privacy settings using tools that most users would otherwise find too difficult to understand and use.

\subsection{Psychological Antecedents of Privacy Decisions}
While machine learning studies have had modest success predicting users' privacy decisions, their results have been scattered; each work only considers a small subset of (one or two) contextual factors, in the context of a single OSN. In this paper, we therefore make an effort to \emph{integrate} new and existing privacy prediction factors into a single comprehensive model. To make this model generalizable across a wide range of OSNs (and, arguably, information systems in general), we theoretically ground this framework in \emph{psychological} research that has identified factors that consistently influence the outcomes of users' privacy decisions.

Existing work has found several psychological factors that influence users' decision making process. For example, Adams identified three major factors that are key to users' privacy perceptions: \emph{information sensitivity}, \emph{recipient} and \emph{usage}~\cite{Adams2000}. These factors are in line with Nissenbaum's theory of \emph{contextual integrity}~\cite{Nissenbaum2009}, which argues that disclosure depends on \emph{context}, \emph{actors}, \emph{attributes}, and \emph{transmission principles} (cf. usage and flow constraints). Based on these theories and other existing research, we identify the following factors:

\begin{itemize} 
	\item \textbf{The user: sharing tendency.} The most widely accepted finding in privacy research is that users differ in their innate tendency to share personal information~\cite{Sheehan2002,Taylor2003,Consolvo2005}. Most prominently, Westin and Harris~\cite{Westin1998} developed a privacy segmentation model which classifies people into three categories: \emph{privacy fundamentalists}, \emph{pragmatists}, and \emph{unconcerned}. Recent work has demonstrated that this categorization might be overly simplistic~\cite{Woodruff2014}. In this light, Knijnenburg et al.~\cite{KnijnenburgJin2013} demonstrated that people's disclosure behavior is in fact multi-dimensional, that is, different people have different tendencies to disclose different types of information (see also~\cite{Olson2005}).
	\item \textbf{The information: sensitivity.} Several studies have found that different types of information have different levels of sensitivity, and that users are less likely to disclose more sensitive information. For instance, Consolvo et al.~\cite{Consolvo2005} and Lederer et al.~\cite{Lederer2003} both found that users are more willing to share vague information about themselves than specific information. Similarly, some privacy decision-making studies found that their manipulation only works for sensitive information~\cite{Acquisti2011}. Note that users occasionally differ in what information they find most sensitive; i.e., Knijnenburg et al.~\cite{KnijnenburgJin2013} find that some users are less willing to publicly share their location than their Facebook posts, while this preference is reversed for others. Generally speaking, though, broad universally applicable levels of information sensitivity can be discerned (e.g. credit card information is much more sensitive than age and gender), and users are typically less willing to share the more sensitive information.
	\item \textbf{The recipient: trustworthiness.} Many studies highlight the trustworthiness of the recipient of the information as an important factor~\cite{Consolvo2005,Hsu2006,Johnson2012,Norberg2007,Olson2005,Toch2010}. Lederer et al.~\cite{Lederer2003} even found that this factor overshadows more traditional contextual factors in terms of determining sharing tendency. Indeed, OSN users tend to restrict access to their profiles by sharing certain information with certain people only~\cite{Kairam2012,Madden2012}, and OSNs have started to accommodate this factor by introducing the facility to categorize recipients into ``groups'' or ``circles''. Research shows that users make extensive use of this facility, albeit often ineffectively~\cite{Kairam2012,Knijnenburg2014,Watson2012}.
	\item \textbf{The context: appropriateness.} Nissenbaum's theory of contextual integrity~\cite{Nissenbaum2009} posits that context-relevant norms play a significant role in users' sharing decisions. Specifically, the theory suggests that disclosure depends on whether it is deemed appropriate or inappropriate in that specific context. Indeed, several scholars argue that the appropriateness of the information request/disclosure plays an important role in determining users' sharing decisions~\cite{Nissenbaum2009,Bansal2008,Borcea2011,Xu2008}. Evaluations of appropriateness are based on users' perception of whether there is a straightforward reason why this recipient should have access to this piece of information in this specific context. This reason is often related to a stated or imagined usage scenario (e.g. I share my work e-mail address [information] with my colleague Dave [recipient] so that he can send me work-related documents [usage]).
\end{itemize}

\section{The Privacy Prediction Model}
The aforementioned psychological antecedents of privacy decisions highlight the incredible complexity of the \emph{privacy calculus}. For each privacy decision, users need to estimate the benefits and risks of disclosure by determining and then integrating all these components: their sharing tendency, the sensitivity of the information, the trustworthiness of the recipient, and the appropriateness of the disclosure in context. Arguably, this is a mentally challenging activity, and it is no surprise that our boundedly rational minds are unable to cope with such complex decisions~\cite{Acquisti2005}.

However, the cited work suggests that these antecedents \emph{do} hold considerable predictive value, meaning that while the privacy calculus may be \emph{mentally} unattainable, it may very well be \emph{computationally} feasible to make consistent predictions of users' preferred privacy decisions based on these antecedents. Indeed, machine learning algorithms are particularly suitable to provide such consistent predictions based on a multitude of anteceding factors. As mentioned, such algorithms have been used in limited cases to predict privacy behaviors~\cite{Ravichandran2009,Sadeh2009,Pallapa2014,Fang2010}. In this section, we will formalize and operationalize this approach. Specifically, we will first provide a generic \emph{formal definition} of the PPM, and discuss a range of algorithms that can be used to implement it. Subsequently, we will operationalize the PPM by proposing an expandable set of \emph{behavioral analogs} of the psychological antecedents, which allow us to unobtrusively measure these antecedents.

\subsection{Theory: A Comprehensive Model Based on Psychological Antecedents}
Formally, PPM models the probability of disclosure $p(D)$ by user $u$ of item $i$ to recipient $r$ in context $c$ as a function of the user's disclosure/sharing tendency $\bar{D_u}$, the sensitivity of the item $S_i$, the trustworthiness of the recipient $T_r$, and the appropriateness of the disclosure in this specific context $A_c$:

\begin{equation}
	p(D_{uirc})=f(\overline{D_u},S_i,T_r,A_c)
\end{equation}

Overly simplistic implementations of this model are the loglinear additive model (Eq. \ref{lin}) and the full factorial loglinear model (Eq. \ref{factor}):

\begin{equation}
	ln(\frac{p(D_{uirc})}{1-p(D_{uirc})})= \alpha + \beta_1 \overline{D_u} + \beta_2 S_i + \beta_3 T_r + \beta_4 A_c
	\label{lin}
\end{equation}

\begin{align}
	ln(\frac{p(D_{uirc})}{1-p(D_{uirc})}) &= \alpha + \beta_1 \overline{D_u} + \beta_2 S_i + \beta_3 T_r + \beta_4 A_c + \beta_5 \overline{D_u} S_i + \beta_6 \overline{D_u} T_r \notag\\
	&+ \beta_7 \overline{D_u} A_c + \beta_8 S_i T_r + \beta_9 S_i A_c + \beta_{10} T_r A_c + \beta_{11} \overline{D_u} S_i T_r \notag\\
	&+ \beta_{12} \overline{D_u} S_i A_c + \beta_{13} \overline{D_u} T_r A_c + \beta_{14} S_i T_r A_c + \beta_{15} \overline{D_u} S_i T_r A_c
	\label{factor}
\end{align}

Machine learning algorithms can provide more sophisticated implementations, such as \emph{decision trees}, \emph{bayesian models}, and \emph{support vector machines}. We will try several of these implementations in our results section. %maybe provide formal definitions of these as well?

\subsection{Practice: Large-scale Prediction Using Behavioral Analogs}

%practice part: how to measure tendency, sensitivity, trustworthiness, and appropriateness? Need to find behavioral analogs
An essential step towards operationalizing PPM is to measure or estimate the parameters $\overline{D_u}$, $S_i$, $T_r$, and $A_c$. This is not a trivial task: these psychological antecedents are hard to quantify, especially on the large scale needed for successful machine learning. Current work defines three ways in which a system can do this:

\begin{itemize}
\item \textbf{Ask the user.} The simplest solution is to directly ask the user, either by having them specify exact values for each parameter, or by allowing them to make a broad classification (e.g. when the system asks the user to categorize a recipient as a ``friend'' or ``acquaintance'', cf.~\cite{Kairam2012,Knijnenburg2014,Watson2012}). This approach is arguably the most accurate (no estimation is involved), but it simply shifts the user's burden from the disclosure decision to the parameter specification (cf. Knijnenburg and Jin \cite{KnijnenburgSigHCI2013} show that this approach does not improve user satisfaction). 
\item \textbf{Derive the values.} Another option is to derive these values from static user variables, such as when the system uses proxy variables, such as when it uses user characteristics (e.g., age, gender, cultural background, or mobile Internet usage) or attitudes (e.g., privacy concerns) to determine users' disclosure profile. This approach is unobtrusive, but rather inaccurate, since these proxy variables are often not context-specific (cf. Knijnenburg et al.~\cite{KnijnenburgJin2013} show that such variables are not very capable of distinguishing between different disclosure profiles). 
\item \textbf{Estimate with available data.} Finally, a system can forego the parameters, and simply try to estimate users' disclosure behaviors based on any available contextual variables. As a finer-grained and thus ultimately more precise method, most existing work uses this approach (cf.~\cite{Ravichandran2009,Sadeh2009,Pallapa2014,Fang2010}). Note, though, that not every available contextual variable has equal predictive value, and using too many contextual variables can lead to overfitting. As such, this approach lacks clear guidance on how to select the most appropriate contextual variables for the estimation process.
\end{itemize}

The PPM approach described in this paper takes a spin on the ``estimate with available data'' method by specifying specific types of data as \emph{behavioral analogs} of the established psychological antecedents. In the following sections, we identify these behavioral analogs in several datasets. Consequently we integrate all factors into a privacy decision making prediction model that can help OSN users to make better privacy decisions.

\section{Data Collection}
Friend requests are the most common and direct way to get access to a user's information in many OSNs. Accepting a friend request discloses at least a part of one's profile and online activities to the requester, so the acceptance or rejection of a friend request is an important privacy decision. In our study, we collected two real-life datasets from Twitter and Goolge+, targeting on the ``Friend requests'' activities to simulate the information disclosure behavior. Besides, location-sharing has gained popularity both in stand-alone apps (e.g. Foursquare, Glympse) and as a feature of existing OSNs (e.g. location-tagging on Facebook and Twitter), which is an information disclosure activity that is particularly strongly influenced by privacy concerns~\cite{Page2012,Zickuhr2012}. Unfortunately, existing location-sharing datasets often do not extend beyond check-in behaviors. We thus created three location-sharing datasets based on a study with manually-collected rich location sharing preferences that includes twenty location semantics, three groups of audiences and several contextual factors such as companion and emotion~\cite{DongJKicwsm}. In the following, we will describe how we collect each dataset and identify the above-mentioned behavioral analogs in these datasets.

\subsection{Twitter Dataset}
Our Twitter dataset consists of a set of Twitter users crawled and classified as legitimate users by Lee et al.~\cite{Lee2011}. We measured the behavioral analogs of our psychological antecedents as follows:

\subsubsection{Disclosure behavior}
In our datasets from both Twitter and Google+, we study users' responses to ``Friend requests'' as our target information disclosure behavior.  Two general friendship mechanism exist on social networks: in \emph{bilateral friendship requests} (e.g. Facebook) friendships are reflexive, and a friendship is only established after the user accepts the request. While this is the ``cleanest'' version of our scenario, the requests themselves are not accessible through the Facebook API, effectively making it impossible to observe rejected requests. Twitter and Google+, on the other hand, use \emph{unilateral friendship requests}: users do not need permission to ``follow'' or ``add to circle'' other users, making one-sided ``friendships'' possible. 

Users who are followed/added to a circle may respond in one of three ways: (1) they may reciprocate the request by following/adding the requester back; (2) they may delete or block the requester; (3) or they may do nothing and simply leave the friendship one-sided. Behavior 1 is observable as a separate, subsequent friendship request; behaviors 2 and 3 are indistinguishable using the Twitter and Google+ APIs. However, since users are notified of being followed, we argue that users will most commonly follow/add the requester back if they accept the request, and otherwise simply delete or block the requester. In other words: behavior 3 occurs rarely for users who are not celebrities (who typically have much more followers than followees). Therefore, our work is based on the assumption that when a user follows the requester back the request is accepted, otherwise it is rejected.

To measure this behavior, we extracted each user $u$'s profile item settings, and following and follower lists. Each friendship request is represented by a tuple $<f(u), f(v), $ $f(u,v), l>$, where $u$ is the requester, $v$ is the receiver and $l$ is the decision label indicating if $v$ accepts (1) or rejects (0) $u$'s request. $f(u)$ and $f(v)$ are collections of features associated with $u$ and $v$ respectively, and $f(u,v)$ represents the relationship between $u$ and $v$.  We classify friend request decisions as follows: 

\begin{definition}[Disclosure behavior] For each friend $u_f$ on $u$'s \emph{following} list, if $u_f$ is also in $u$'s \emph{followers} list, that is, if $u_f$ also follows $u$, we say that $u_f$ accepted $u$'s request, otherwise $u_f$ rejected $u$'s request. 
\end{definition}
	
Consequently, we extracted $u_f$'s profile items and following and follower lists as well. This results in a total of 17,118 users in our dataset. We removed ``verified'' users, since most of such users are celebrities who often have many more followers than followees; our definition of ``accepting/rejecting a friend request'' will arguably not hold for such users.

Note that based on our Twitter dataset we are unable to distinguish who follows whom first; when two users $u_1$ and $u_2$ follow each other, this results in two records $<f(u_1), f(u_2),$ $f(u_1,u_2), 1>$ and $<f(u_2), f(u_1), f(u_2,u_1), 1>$. Only one of these describes the actual reciprocation of a friend request, the other is spurious. In the Twitter dataset this results in a noisy dataset where only half of the signals are real. In the Google+ dataset we present below, we are able to untangle the chronological order of friend requests. This dataset thus arguably provides more accurate results.

\subsubsection{Sharing tendency of the information holder}
\emph{FollowTendency} is a behavioral analog of users' \emph{disclosure/sharing tendency}, defined as the relative number of people they follow. Formally speaking:
\begin{definition}[Sharing tendency] $followTendency = \frac{\#following}{\#follower  + \#following} $
\end{definition}

\subsubsection{Trustworthiness of the recipient}
Our behavioral analog of the recipient's \emph{trustworthiness} is based on the intuition that a user with relatively many followers is likely to have a higher reputation, and thus more trustworthy. Formally speaking:
\begin{definition}[Trustworthiness] $trustworthiness = \frac{\#follower}{\#follower  + \#following}$
\end{definition}
 
\subsubsection{Sensitivity of the requested information}
We argue that the sensitivity $S_i$ of a Twitter user $u$'s profile item $i$ depends on how common the user's value of $i$ is in the population: the more common the value, the less sensitive the information. Formally speaking: 
\begin{definition}[Sensitivity]\label{def:sensitivity} Suppose a profile item $i$ has $m$ possible settings $\{i_1, i_2, \ldots, i_m\}$ ($m \ge 1$). The distribution of different settings over the whole population is $P^{i} = \{p_{i_1}, p_{i_2}, \ldots, p_{i_m}\}$, where $0 \le p_{i_j} \le 1$ and $\sum^{m}_{j=1} p_{i_j} = 1$. If user $u$ sets his / her profile item $i$ as $i_k$ ($1 \le k \le m$), the sensitivity value of $S_i = \frac{1}{p_{i_k}}$.
\end{definition}

On Twitter, users have the option to set profile items \emph{URL}, \emph{GEO} and \emph{Protected}. \emph{GEO} and \emph{Protected} are boolean values indicating whether the user has enabled the automatic geo-tagging of her tweets, and whether her profile is protected (i.e. her updates can only be followed with explicit consent from the user). \emph{URL} is a field that users can use to enter a personal website. We categorize its value as either \emph{blank}, a \emph{personal} URL (linking to Facebook or LinkedIn), or an \emph{other} URL. We calculate the corresponding sensitivity scores for these items and use them as our behavioral antecendents of \emph{sensitivity}. 

\subsubsection{Appropriateness of the request}
Friendship requests are more appropriate if there is a lot of existing overlap between the two users' networks. Formally speaking:
\begin{definition}[Appropriateness] The approrpiateness of a friend request of user $v$ to user $u$ depends on the overlap between their networks, which can be measured with the following indicators:
	\begin{itemize}
    	 \item $JaccardFollowing_{(u,v)} = \frac{ |followings(u) \cap followings(v)| }{|followings(u)  \cup followings(v)|}$ 
     	 \item $JaccardFollower_{(u,v)} = \frac{ |followers(u) \cap followers(v)| }{|followers(u) \cup followers(v)|}$
		 \item $comFollowing(u) = \frac{\#commonFollowing}{\#following(u)}$
		 \item $comFollower(u) = \frac{\#commonFollower}{\#follower(u)}$
    	 \item $comFollowing(v) = \frac{\#commonFollowing}{\#following(v)}$
		 \item $comFollower(v) = \frac{\#commonFollower}{\#follower(v)}$
	\end{itemize}	 
\end{definition}

\subsection{Google+ Dataset}
Gong et al.~\cite{Gong2011} crawled the whole evolution process of Google+, from its initial launch to public release. The dataset consists of 79 network snapshots, these stages can be used to uncover a rough  chronological account of friendship creation (i.e. users adding each other to their circles). We focus on the first two stages to build our dataset, where ``who sends the friend request to whom first'' can be identified by the stage ids. In total, the set contains 3,481,544 active users\footnote{We call a user with at least one following as an ``active user''.} in stage 0 and 14,289,211 in stage 1. 

We used the same factors in our Twitter dataset to measure the behavioral analogs of our psychological antecedents. Specifically, we use the ``add to circle'' activity to simulate the information {\bf disclosure behavior}. Each instance in our Google+ dataset is a tuple $<f(u), f(v), f(u,v), l>$, where $l$ indicates $v$' decision to reciprocate $u$'s request (1) or not (0). 

We use the same behavioral analogs as those defined in Twitter dataset for {\bf sharing tendency}, {\bf trustworthiness}, {\bf sensitivity}, and the {\bf appropriateness} of the request. For {\it sensitivity},  we use profile attributes available on Google+, namely \emph{Employer}, \emph{Major}, \emph{School} and \emph{Places} that are either publicly displayed (1) or not (0). We thus calculate the following four behavioral analogs for \emph{sensitivity}: (1) \emph{S(Employer)}, (2) \emph{S(Major)}, (3) \emph{S(School)} and (4) \emph{S(Places)}. 

\subsection{Location Sharing Datasets}
Our location sharing datasets are built based on a survey on location sharing preferences. We conducted this survey by recruiting 1,088 participants using Amazon Mechanical Turk\footnote{https://www.mturk.com/mturk/}. We restricted participation to US Turk workers with a high worker reputation who had previously used a form of location sharing services. The demographics are shown in Table~\ref{tab:locAtt}.

\begin{table}[h]
  \centering
  \begin{tabular}{|c|c|}
    \hline
    \textbf{Attribute} & \textbf{Distribution}  \\ \hline \hline
    \multirow{3}{*}{Age} &18 to 24 (21.42\%), 25 to 34 (45.23\%), \\ 
    & 35 to 44 (20.23\%), 45 to 54 (5.95\%) \\
    & 55 to 64 (7.14\%)  \\  \hline
    Gender & male (57.14\%), female (42.85\%)  \\ \hline
    Marriage & married (40.47\%), not married (59.52\%) \\ 
    \hline
  \end{tabular}
  \caption{Demographics of the participants in the location sharing preference study.}
  \label{tab:locAtt}
\end{table}

We specifically asked the participants about their privacy concern in the location sharing survey. The distribution of their claimed {\it privacyLevel} is: {\it Very Concerned} (39\%), {\it Moderately} (41\%), {\it Slightly} (15\%), {\it Not Care} (5\%). Consistent with previous research,
as many as 80\% of the participants claimed to be moderately or very concerned about their privacy~\cite{Westin1998,HoMA09}.

\subsubsection{Disclosure behavior}
We constructed our dataset by requesting users' feedback to systematically manipulated location sharing scenarios. In each scenario, participants were asked to indicate whether they would share their location with three different types of audience: \emph{Family}, \emph{Friend} and \emph{Colleague}.

We ran 5 different studies to collect our data: 
\begin{itemize}
	\item In study 1, each scenario consisted of one of twenty location semantics supported by Google Places\footnote{https://developers.google.com/places/}: \emph{Airport, Art Gallery, Bank, Bar, Bus Station, Casino, Cemetery, Church, Company Building, Convention Center, Hospital, Hotel, Law Firm, Library, Movie Theater, Police Station, Restaurant, Shopping Mall, Spa} and \emph{Workplace}. 
	\item In study 2, each scenario consisted of a location, plus a certain \emph{time}: on a weekday during the day, on a weekday at night, on the weekend. %is this correct?
	\item In study 3, each scenario consisted of a location, plus a \emph{companion}: alone, with family, with friends, or with colleagues.  % when I did the experiment, I further classified the original 7 categories of companion into four. 
	\item In study 4, we combined location, time and companion in each scenario.
	\item In study 5, each scenario consisted of a location, plus an \emph{emotion}: positive or negative. 
	\end{itemize}
Some of these contextual variables match features available on location sharing services, such as ``who are you with'', and emotion icons on Facebook.  In total, five different scenarios were presented to the user, as shown in Table~\ref{tab:LocSetting}. Each participant was randomly assigned ten scenarios with different combinations of location and contextual information. 

\begin{table}[h]
  \centering
  \begin{tabular}{|c|c|c|c|}
    \hline
    Study & Scenario & \#Participants &\#Records  \\ \hline \hline   
    1& location & 84 & 840 \\ 
    2& location+time & 133 & 1376 \\
    3 & location+ companion & 244 & 2440 \\
    4 & location + time + companion & 510 & 4969 \\
    5 & location + emotion & 117 & 1170 \\ \hline
     \end{tabular}
  \caption{Experimental settings of location sharing preference study.}
  \label{tab:LocSetting}
\end{table}

For each targeted group of audience $V$ (\emph{Family}, \emph{Friend} and \emph{Colleague}), we collect the associated sharing records represented by tuples $<f(u), f(u,V, loc), f(loc), l(V)>$, which results in 3 location sharing datasets. $f(u)$ represents the user's features. $f(u,V, loc)$ describes the relationship between the three parties, i.e., \emph{user}, \emph{audience} and \emph{location}.  As the sharing information is the given location, we specifically include the features $f(loc)$ with regard to the current location $loc$ into each tuple. $l(V)$ is the decision label indicating if $u$ shares her location with audience $V$ (1) or not (0). 

% including demographics such as \emph{gender},  \emph{age},  \emph{marriage} and the claimed \emph{privacyLevel}. The values to each of these attributes $a$ are all categorical values, denoted as $c(a)$. 
\subsubsection{Sharing tendency of the information holder} 
In this ``voluntarily information sharing'' scenario, sharing tendency of the information holder largely influences his/her privacy decisions. In our location sharing preference study, each user $u$ only has one sharing option to each group of audiences under a specific scenario. That is, we do not have the ``historical'' sharing records of $u$ with the same scenario to predict the current or future privacy decision. Therefore, we choose to use other users' sharing behavior to estimate individual $u$'s sharing probability.  

One type of estimation on sharing tendency is based on a specific feature $Q$ over all the populations in the dataset, i.e., {\it overall sharing probability}, represented by $p^u(Q)$. That is, $p^u(Q)$ represents the sharing tendency of $u$ based on feature $Q$. We estimate it by the sharing probability of the users in the given dataset R who have the same feature value of $Q$ with $u$, regardless of other information such as sharing audience, current location, etc. Formally speaking, 
\begin{definition}[overall sharing probability]\label{def:overallSharing} Suppose a feature $Q$ has $m$ possible values $\{q_1, q_2, \ldots, q_m\}$ ($m \ge 1$). The sharing probability of the users with different feature values on $Q$ over the whole records $R$ is $P^u(Q) = \{p_{q_1}, p_{q_2}, \ldots, p_{q_m}\}$, where $0 \le p_{q_i} \le 1$ and $\sum^{m}_{i=1} p_{q_i} = 1$. That is, $p_{q_i}$ represents the sharing probability of the users with $q_i$ as the value of feature $Q$. If user $u$'s feature value on $Q$ is $q_k$ ($1 \le k \le m$), the overall sharing probability of $u$ based on feature $Q$ is $p^u(Q) = p_{q_k}$.
\end{definition}

In this study, we consider the \emph{disclosure tendency} of $u$ from the four aspects: demographic features \emph{age}, \emph{gender} and \emph{marriage}, as well as the claimed \emph{privacyLevel}. It results in four types of overall sharing probability, i.e., $p^u(age)$, $p^u(gender)$, $p^u(marriage)$ and $p^u(privacyLevel)$.

Besides the specific feature $Q$, the other type of estimation on sharing tendency also considers the contextual variable $\alpha$. We call such estimation as \emph{$\alpha$-conditional sharing probability} and denoted it as $p^u_{\alpha}(Q)$. We use the sharing probability of the users in R with the same attribute value on $Q$ who have been under the same scenario $\alpha$ to estimate $u$'s sharing probability. Formally speaking, 
\begin{definition}[$\alpha$-conditional sharing probability]\label{def:alphaSharing} Suppose an attribute $Q$ has $m$ possible values $\{q_1, q_2, \ldots, q_m\}$ ($m \ge 1$). The sharing probability of the users with different attribute values on $Q$ over the whole set of sharing records $R^{\alpha}$ under scenario $\alpha$ is $P^u_{\alpha}(Q)  = \{p^{\alpha}_{q_1}, p^{\alpha}_{q_2}, \ldots, p^{\alpha}_{q_m}\}$, where $0 \le p^{\alpha}_{q_i} \le 1$ and $\sum^{m}_{i=1} p^{\alpha}_{q_i} = 1$. If user $u$'s feature value on scenario $Q$ is $q_k$ ($1 \le k \le m$), the $\alpha$-conditional sharing probability of $u$ based on $Q$ under $\alpha$ is $p^u_{\alpha}(Q) = p^{\alpha}_{q_k}$.
\end{definition}

In this study, we set $\alpha$ as the current location $loc$ or the audience $V$, and consider companion, emotion, time as well as the $loc$ and $V$ as possible contextual variables. That is, we will consider the following $\alpha$-conditional sharing probabilities:  $p^u_{loc}(companion)$,  $p^u_{loc}(emotion)$, $p^u_{loc}(time)$, $p^u_{V}(companion)$,  $p^u_{V}(emotion)$,  $p^u_{V}(time)$, $p^u_{loc}(V)$ and $p^u_{V}(loc)$.

When building the PPM, we choose the corresponding {\it overall sharing probability} and {\it $\alpha$-conditional sharing probability} to represent the {\it disclosure/sharing tendency} $\overline{D_u}$ according to the different scenarios listed in Table~\ref{tab:LocSetting}.

\subsubsection{Trustworthiness of the recipient}
As location sharing is a voluntarily sharing activity, the information holder usually makes the privacy decisions partially based on the trustworthiness of the audience. Typically, the higher of the probability $u$ is willing to share to the given type of audience $V$, the higher trustworthiness of $V$ is believed by $u$. According to the above definitions on sharing tendency, we can formally define trustworthiness as follows: 
\begin{definition}[Trustworthiness] The trustworthiness of $u$ to the sharing audience $V$ is estimated by the {\it audience-conditional sharing probability} of $u$ under the contextual variable $V$ without consider other features. That is, $trustworthiness(V) = p^u_{V}(\cdot)$.
\end{definition}

\subsubsection{Sensitivity of the shared location}
The sensitivity of the location being shared is of vital importance to the privacy decisions. Usually, the more people is willing to share a location $loc$, the less they think the location is sensitive. Thus, we can use the  {\it location-conditional sharing probability} regardless of any features to estimate the sensitivity of the shared location. Formally speaking, 
\begin{definition}[Sensitivity] The sensitivity of a given $loc$ being shared by $u$ is defined as: $sensitivity(loc) = p^u_{loc}(\cdot)$.
\end{definition}
 
\subsection{Final datasets}
The final datasets used in our study is shown in Table~\ref{tab:datasets}, including the basic statistics on the final privacy decisions. 
\begin{table}[h]
  \centering
  \begin{tabular}{|c|c|}
    \hline 
     Dataset  & Statistics \\ \hline
    \multirow{2}{*}{\textbf{Twitter Dataset}}   &  $D_{Twitter} =TSet_{req} \cup TSet_{rec}$   \\ 
     				                                   & (\#{\it accepted}: 4,874; \#{\it rejected}: 7,914)  \\  \hline
    \multirow{2}{*}{\textbf{Google+ Dataset}}  &  $D_{Google+} = GSet_{req} \cup GSet_{rec}$ \\  
                                                                      &  (\#{\it accepted}: 21,798; \#{\it rejected}: 114,400) \\  \hline

    \multirow{3}{*}{\textbf{Location Dataset}}  &  $D_{Family}$: \#{\it shared}: 8,241; \#{\it not shared}: 2,554 \\  \cline{2-2} 
								   &  $D_{Friend}$: \#{\it shared}: 8,030; \#{\it not shared}: 2,845 \\   \cline{2-2} 
 								   &  $D_{Colleague}$: \#{\it shared}: 5,249; \#{\it not shared}: 5,546  \\  \hline
  \end{tabular}
  \caption{Statistics of datasets.}
  \label{tab:datasets}
\end{table}
 
\section{PPM: Privacy Prediction Model}
Before building the privacy prediction model, we have analyzed and proved the above defined behavioral analogs work rather well on the all the datasets listed in Table~~\ref{tab:datasets}~\cite{DongJKicwsm}. 

In this section we talk about how we build the PPM using the above presented behavioral analogs of the following psychological antecedents:
\begin{itemize}
 \itemsep0em
\item (1) \emph{Trustworthiness} of the requester/ information audience 
\item (2) \emph{Sharing tendency} of the receiver/ information holder
\item (3) \emph{Sensitivity} of the requested/ shared information
\item (4) \emph{Appropriateness} of the request 
\end{itemize}

Note that we also include:
\begin{itemize}
 \itemsep0em
\item (5) \emph {Other contextual factors}
\end{itemize}

The PPM is aimed to help OSN users to manage their privacy by predicting their disclosure behavior and recommending privacy settings in line with this behavior. Specifically, based on these behavioral analogs of the psychological antecedents, we build a binary classification model that learns the influence of these features on privacy decisions (i.e. sharing/disclosure decisions, such as \emph{accept} vs. \emph{reject}, or \emph{share} vs. \emph{not share}). 

\subsection{Machine learning outcomes}

We build a decision making model for each of the five datasets described in Table~\ref{tab:datasets}. One problem with the five datasets is they are imbalanced, that is, the number of accepts/shares is much larger or smaller than the number of rejects/not-shares. We employ the common machine learning practice -- {\it undersampling}, to balance the sets by randomly sampling items from the ``large class'' to match the size of the ``small class''. 
We use an adapted {\it 10-fold cross validation} approach (detailed implementation is described in the discussion in Section~\ref{sec:discuss}) to split the training and testing datasets. The final results are averaged over the classification results in all the folds. The commonly used \emph{F1} and \emph{AUC} are employed as evaluation metrics. \emph{F1} is the harmonic mean of precision and recall, and \emph{AUC} is a statistic that captures the precision of the model in terms of the tradeoff between false positives and false negatives. The higher of these values, the better of the performance. 
We use several of the binary classification algorithms provided by Weka~\cite{weka2009} to build our models, including {\it J48}, {\it Na\"{\i}ve Bayes}, {\it Support Vector Machine (SVM)}, etc. Among them, \emph{J48} produced the best results in terms of both {\it F1} and {\it AUC}. The corresponding results are shown in Table~\ref{tab:result}. As seen, our privacy decision making prediction model has a good performance (cf.~\cite{Swets1988}).

\begin{table}[h]
  \centering
  \begin{tabular}{|c|c|c|c|}
    \hline 
    \multirow{2}{*}{Dataset} & \multirow{2}{*}{\#tuples} &\multicolumn{2}{|c|} {\textbf{Measurements}} \\ \cline{3-4}
    & &F1 & AUC \\ \hline \hline
    $D_{Twitter}$ &  9,748 & 0.796   &  0.85\\  \hline
    $D_{Google+}$ & 43,596  &  0.898    &  0.899 \\  \hline
   $D_{Family}$  &5,378  & 0.845 & 0.879\\  \hline
    $D_{Friend}$  & 5,822  & 0.810 & 0.844 \\  \hline
    $D_{Colleague}$  &11,054  & 0.737 & 0.752\\  \hline
  \end{tabular}
  \caption{Results of privacy decision making prediction model.}
  \label{tab:result}
\end{table}

\begin{table*}[t]
  \centering
  \begin{tabular}{|c|c|c|c|c|c|c|}
    \hline 
     \multirow{3}{*}{Dataset} & \multicolumn{6}{|c|}{F1} \\ \cline{2-7}
     & \multirow{2}{*}{All} & \multicolumn{5}{|c|}{removed features} \\ \cline{3-7}
     &  & (1) & (2)  & (3) & (4)  & (5) \\ \hline
    $D_{Twitter}$   &  0.796 & 0.784  & \textbf{0.751} & 0.796 & \textbf{0.767} & -  \\  \hline
    $D_{Google+}$ & 0.898  & 0.889 & \textbf{0.887} &  0.898  &  0.889 & - \\  \hline
    $D_{Family}$ & 0.845 &- &  0.840 &  \textbf{0.833} &  \multicolumn{2}{|c|}{\textbf{0.833}}  \\  \hline
    $D_{Friend}$ & 0.810 & - &  \textbf{0.798} & 0.802 &   \multicolumn{2}{|c|}{ 0.800} \\  \hline
    $D_{Colleague}$  & 0.737 & - &  0.730  &  \textbf{0.726} & \multicolumn{2}{|c|}{ 0.727} \\  \hline \hline
    
     \multirow{3}{*}{Dataset} & \multicolumn{6}{|c|}{AUC} \\ \cline{2-7}
     & \multirow{2}{*}{All} & \multicolumn{5}{|c|}{removed features} \\ \cline{3-7}
     &  & (1) & (2)  & (3) & (4)  & (5) \\ \hline
    $D_{Twitter}$   &  0.850 & 0.835 & \textbf{0.812} & 0.850 & \textbf{0.832} &- \\  \hline
    $D_{Google+}$ & 0.899   & 0.892 & \textbf{0.891}  & 0.898 & \textbf{0.890}  & -  \\  \hline
    $D_{Family}$ & 0.879  & -  & 0.875  & \textbf{0.867}  & \multicolumn{2}{|c|}{0.870}  \\  \hline
    $D_{Friend}$ & 0.844 & -  & 0.840 & 0.839 &\multicolumn{2}{|c|}{\textbf{0.835}} \\  \hline
    $D_{Colleague}$  & 0.752 & -  & 0.748 &  \textbf{0.743} &\multicolumn{2}{|c|}{0.745} \\  \hline \hline
    
  \end{tabular}
  \caption{Performance of using different feature sets in decision making prediction model.}
  \label{tab:factorResult}
\end{table*}

We further verify the effectiveness of each factor by testing the privacy decision making model without the corresponding factor. Specifically, Table~\ref{tab:factorResult} compares the performance of the decision making models with all features (\emph{All}) against their performance after removing the features belonging to each of the factors using {\it J48}: (1) \emph{trustworthiness}; (2) \emph{sharing tendency}; (3) \emph{sensitivity}; (4) \emph{appropriateness}; (5) \emph{contextual factors}\footnote{As contextual factors are not studied in the \emph{friend requests} scenario, we have no results for removing such factors from $D_{Twitter}$ and $D_{Google+}$. Similarly, although \emph{trustworthiness} of the audience is an important factor in the location sharing study, our privacy decision making model is built for audience separately (as these measures are repeated per scenario). Thus, \emph{trustworthiness} is not a feature in $D_{Family}$, $D_{Friend}$ and $D_{Colleague}$. Finally, the features belonging to \emph{appropriateness} are difficult to split off from \emph{contextual factors} in the location study, so those results are combined.}.

Table~\ref{tab:factorResult} shows that removing some factors may reduce the prediction performance. In line with our feature ranking results in~\cite{DongJKicwsm}, this is mainly true for the \emph{trustworthiness} of requester and the \emph{follow tendency} of the receiver as well as the \emph{appropriateness} of the request in the \emph{friend requests} scenario of Twitter and Google+. These factors are more important than the \emph{sensitivity} factors, probably because we simply use user profile to quantify it. In the \emph{location sharing} scenario, the \emph{sharing tendency} of the user, the \emph{sensitivity} of the location as well as the \emph{contextual factors} are all important predictors that reduce the \emph{F1} and \emph{AUC} values when excluded. For different types of recipients, the dominate factors to privacy decisions vary as well. For instance, when the recipients/audiences are colleagues and family members, the sensitivity of the shared location is more important. However, the information holder's sharing tendency has a bigger influence on the privacy decisions when the recipients are their friends.  

\subsection{Discussion}\label{sec:discuss}
\subsubsection{Class imbalance problem}
As we mentioned earlier, class imbalance is a common but serious problem when building classification models. To deal with imbalanced datasets, {\it oversampling} and {\it undersampling} are the two most commonly used techniques that use a bias to select more samples from one class than from another in order to balance the dataset. Specifically, {\it oversampling} is to repeatedly draw samples from the smaller class until the two classes have the same size. This technique may results in biased distribution due to certain over-repeated samples from the smaller class. While, {\it undersampling} is to randomly select the same number of instances from the bigger class as the number of instances in the smaller class. As some instances in the bigger class are not included in the final dataset, it may lose some information and result in a biased distribution. In our approach, we choose to use {\it undersampling} because of another approach adopted -- {\it adapted 10 fold  cross validation}.  Traditional {\it k-fold cross validation} splits the same balanced dataset into $k$ parts, then each time uses 1 part of the dataset as testing dataset and the remaining $k$-1 parts as training dataset to evaluate the performance. This process repeats $k$ times with different part selected as testing dataset and the final results are averaged over the $k$ folds. In our work, instead of using the same balanced dataset and repeats 10 times, we built a different balanced dataset using {\it undersampling} each time and applied {\it 10-fold cross validation}. In this way, all the balanced datasets together may cover all or the majority of the instances in the bigger class. The results are actually averaged over the 100 different training-testing datasets, which can largely reduce the potential bias caused by {\it undersampling}.

\subsubsection{Missing values in location datasets}
There are a few differences when we built PPM for Google+/Twitter datasets and the three location datasets.  In Google+ and Twitter datasets, each instance has the same number of attributes, represented by the measurements we defined for the behavioral analogs of those psychological antecedents, including {\it trustworthiness}, {\it sharing tendency}, {\it sensitivity} and {\it appropriateness}. Thus, any machine learning algorithms can be directly applied on the datasets. For the three location datasets, however, the {\it disclosure/sharing tendency} will be measured by the {\it overall sharing probability} or the {\it $\alpha$-conditional sharing probability}, which is based on the contextual variables, e.g. {\it companion}, {\it time}, {\it emotion}. As described in Table~\ref{tab:LocSetting}, the five location-sharing scenarios we studied used different subsets of these contextual variables. By putting the data from these fives studies together, we thus create missing values for some of the attributes. We deal with this problem using the embedded {\it ReplaceMissingValues} filter in Weka. That is, Weka will deal with the missing values using different mechanisms given the different machine learning algorithms. Internally, {\it J48} replaces the missing values based on the weighted average value proportional to the frequencies of the non-missing values. As {\it J48} produces the best results not only on all the three location datasets, but also on Google+ and Twitter datasets, we may claim that the way missing values are solved by {\it 48} is acceptable and may be a preferable approach compared with other possible approaches. Besides, the best performance of PPM achieved by {\it J48} does not purely rely on its mechanism of dealing with missing values.

\subsubsection{PPM on simulation datasets}
Table~\ref{tab:result} shows that PPM produces better results on the $D_{Google+}$ dataset than on the $D_{Twitter}$ dataset in terms of both {\it F1} score and {\it AUC}. As we mentioned earlier, the Google+ dataset simulates the ``friend request'' scenario more accurately than the Twitter dataset, as the chronological order of friendship invitations is captured in our Google+ dataset but not in the Twitter dataset. The better performance of PPM on the more accurate dataset arguably demonstrates the effectiveness of the proposed privacy prediction model.

\section{Limitations}
In our studies, we demonstrated that the factors identified in the PPM are essential in \emph{predicting} users' disclosure/sharing behavior. The ultimate goal of the PPM is to give personalized privacy decision support to overwhelmed OSN users, and our first limitation is that we did not actually implement such a decision support system. We plan to do this in our future work, and point to~\cite{Knijnenburg2015} for initial results in the context of recommender systems rather than OSNs. Aside from this, there are some other limitations with regard to our study that can be accounted for in future work. 

The first limitation is our datasets: users' real sharing behavior data (rather than questionnaire data) is more accurate in identifying true factors that affect user decision making. We make use of a large dataset of Twitter and Google+ users to analyze such real behavior, but we use an imperfect proxy of users' actual ``friend requests".  Moreover, the location sharing data is based on imagined scenarios rather than real sharing behavior. This work takes a first step towards comprehensively and accurately study the factors affecting users' privacy decision making; in the future we hope to study these factors using real behavioral data such as true friend request acceptance and location sharing behaviors on social network sites.

The second limitation involves the identified factors. We analyzed the most important factors in privacy decision making as identified by existing work; but the study of factors determining privacy decisions has been far from comprehensive in the past. To create a more generally applicable model, we also specifically selected factors that are applicable to a wide range of privacy decision making scenarios in different social media, not restricted to online social networks. Consequently, we may miss system/domain-specific factors that could have a significant influence on privacy decision making. Finally, we tried to focus on very simple behavioral analogs in our study; more sophisticated analogs could likely be identified that further increase the prediction performance. %Furthermore, we did not analyze or quantify the importance of each factor compared to the other factors. So we can not compare our findings with the finding in other research such as~\cite{Lederer2003}.

Our third limitation is that our privacy decision making prediction model is built as a binary classifier. In practical applications of, say, a ``privacy recommender'', it is not always appropriate to make ``hard'' recommendations to users, e.g., to pervasively recommend either ``accept'' or ``reject''. Instead, it might be more applicable to calculate a ``privacy risk'' score based on the identified factors and let user to make the final decision based on the score, or to only intervene when the calculated risk passes a certain (user-defined) threshold. 

\section{Design implications and Conclusion}
In this paper, we developed a generic Privacy Prediction Model based on known psychological antecedents of privacy decision making. We operationalized the PPM by identifying behavioral antecedents of the psychological factors, and analyzed how these factors influenced privacy decisions in several real-world and collected datasets. While our work does not move beyond prediction, it proves the feasibility of an automated tool to predict users' disclosure/sharing behavior. Such a tool can help ease the burden on the user, who on most OSNs has to make an almost unreasonable number of privacy decisions. Our investigation specifically led to the following important observations that may affect future design of OSNs: 
\begin{itemize}
\item Consistent with previous studies in \emph{information sharing}, \emph{who} is the information audience is an important factor deciding if the information will be shared. Similarly, in the scenario of \emph{information requests}, \emph{who} is the requester determines the \emph{trust} from the receiver, therefore determines the final privacy decision making outcome. 
\item As self-representation is one of the main purposes of OSNs, users' privacy decision making does not exclusively depend on their privacy concern, but more generally on the tradeoff between privacy and self-presentation. We have captured this in the definition of \emph{sharing tendency}.
\item \emph{Sensitivity} is not an objective concept; it varies with audience. For example, in our location sharing preference study, locations such as \emph{Bar}, \emph{Casino} are more sensitive if shared to \emph{Colleague} or \emph{Family} compared to \emph{Friend}. This effect may be captured by the \emph{appropriateness} of the request. 
\item Although the assumption of ``rationality'' in privacy decision making has been criticized by many studies, users do consider the \emph{appropriateness} of the request/sharing activity. Consistent with our intuition and previous studies, users tend to share information with or accept requests when this is appropriate in the current context. 
\item \emph{Contextual information} is an indispensable factor in privacy decision making, primarily due to its effect on \emph{appropriateness}. However, the effect of different contextual factors varies.
\end{itemize}
 
Concluding, privacy decision making is a trade-off between the potential benefit and risk; a tradeoff that is rather difficult for users to make. Our privacy decision making prediction model combines several important psychological and contextual factors that influence this tradeoff, and learns their functionality by building a binary classifier. The proposed privacy decision making prediction model produces good results based on the five identified factors, and can be used to assist users to protect their privacy in online social networks.

% Appendix
%\appendix
%\section*{APPENDIX}
%\setcounter{section}{1}
%In this appendix, we measure
%the channel switching time of Micaz [CROSSBOW] sensor devices.
%In our experiments, one mote alternatingly switches between Channels
%11 and 12. Every time after the node switches to a channel, it sends
%out a packet immediately and then changes to a new channel as soon
%as the transmission is finished. We measure the
%number of packets the test mote can send in 10 seconds, denoted as
%$N_{1}$. In contrast, we also measure the same value of the test
%mote without switching channels, denoted as $N_{2}$. We calculate
%the channel-switching time $s$ as
%\begin{eqnarray}%
%s=\frac{10}{N_{1}}-\frac{10}{N_{2}}. \nonumber
%\end{eqnarray}%
%By repeating the experiments 100 times, we get the average
%channel-switching time of Micaz motes: 24.3$\mu$s.
%
%\appendixhead{ZHOU}
%
%% Acknowledgments
%\begin{acks}
%The authors would like to thank Dr. Maura Turolla of Telecom
%Italia for providing specifications about the application scenario.
%\end{acks}

% Bibliography
\bibliographystyle{ACM-Reference-Format-Journals}
\bibliography{PPM_privacyPrediction}


\begin{thebibliography}{00}

%%% the Web address.

\ifx \showCODEN    \undefined \def \showCODEN     #1{\unskip}     \fi
\ifx \showDOI      \undefined \def \showDOI       #1{{\tt DOI:}\penalty0{#1}\ }
  \fi
\ifx \showISBNx    \undefined \def \showISBNx     #1{\unskip}     \fi
\ifx \showISBNxiii \undefined \def \showISBNxiii  #1{\unskip}     \fi
\ifx \showISSN     \undefined \def \showISSN      #1{\unskip}     \fi
\ifx \showLCCN     \undefined \def \showLCCN      #1{\unskip}     \fi
\ifx \shownote     \undefined \def \shownote      #1{#1}          \fi
\ifx \showarticletitle \undefined \def \showarticletitle #1{#1}   \fi
\ifx \showURL      \undefined \def \showURL       #1{#1}          \fi

\bibitem[\protect\citeauthoryear{Acquisti and Gross}{Acquisti and
  Gross}{2006}]%
        {Acquisti06}
{A Acquisti} {and} {Ralph Gross}. 2006.
\newblock \showarticletitle{Imagined communities: Awareness, information
  sharing, and privacy on the Facebook}. In {\em In 6th Workshop on Privacy
  Enhancing Technologies}. 36--58.
\newblock


\bibitem[\protect\citeauthoryear{Acquisti and Grossklags}{Acquisti and
  Grossklags}{2005}]%
        {Acquisti2005}
{Alessandro Acquisti} {and} {Jens Grossklags}. 2005.
\newblock \showarticletitle{Privacy and rationality in individual decision
  making}.
\newblock {\em IEEE Security \& Privacy\/}  {2} (2005), 24--30.
\newblock


\bibitem[\protect\citeauthoryear{Acquisti, John, and Loewenstein}{Acquisti
  et~al\mbox{.}}{2012}]%
        {Acquisti2011}
{Alessandro Acquisti}, {Leslie~K John}, {and} {George Loewenstein}. 2012.
\newblock \showarticletitle{The {Impact} of {Relative} {Standards} on the
  {Propensity} to {Disclose}}.
\newblock {\em Journal of Marketing Research\/} {49}, 2 (2012), 160--174.
\newblock
\showISSN{0022-2437, 1547-7193}
\showDOI{%
\url{http://dx.doi.org/10.1509/jmr.09.0215}}


\bibitem[\protect\citeauthoryear{Adams}{Adams}{2000}]%
        {Adams2000}
{Anne Adams}. 2000.
\newblock \showarticletitle{Multimedia Information Changes the Whole Privacy
  Ballgame}. In {\em Proceedings of the Tenth Conference on Computers, Freedom
  and Privacy: Challenging the Assumptions} {\em (CFP '00)}. ACM, 25--32.
\newblock
\showISBNx{1-58113-256-5}


\bibitem[\protect\citeauthoryear{Al~Hasib}{Al~Hasib}{2009}]%
        {AAHasib2009}
{Abdullah Al~Hasib}. 2009.
\newblock \showarticletitle{Threats of online social networks}.
\newblock {\em IJCSNS International Journal of Computer Science and Network
  Security\/} {9}, 11 (2009), 288--293.
\newblock


\bibitem[\protect\citeauthoryear{Baker, Buoni, Fee, and Vitale}{Baker
  et~al\mbox{.}}{2011}]%
        {Baker2011}
{Douglas Baker}, {Nicole Buoni}, {Michael Fee}, {and} {Caroline Vitale}. 2011.
\newblock \showarticletitle{Social networking and its effects on companies and
  their employees}.
\newblock {\em Retrieved November\/}  {15} (2011).
\newblock


\bibitem[\protect\citeauthoryear{Balebako, Leon, Mugan, Acquisti, Cranor, and
  Sadeh}{Balebako et~al\mbox{.}}{2011}]%
        {balebako2011}
{R. Balebako}, {P.~G. Leon}, {J. Mugan}, {A. Acquisti}, {L.~F. Cranor}, {and}
  {N. Sadeh}. 2011.
\newblock \showarticletitle{Nudging users towards privacy on mobile devices}.
  In {\em {CHI} 2011 workshop on {Persuasion}, {Influence}, {Nudge} and
  {Coercion} {Through} {Mobile} {Devices}}. Vancouver, Canada, 23--26.
\newblock
\showURL{%
\url{http://www.andrew.cmu.edu/user/jmugan/Publications/chiworkshop.pdf}}


\bibitem[\protect\citeauthoryear{Bansal, Zahedi, and Gefen}{Bansal
  et~al\mbox{.}}{2008}]%
        {Bansal2008}
{Gaurav Bansal}, {Fatemeh Zahedi}, {and} {David Gefen}. 2008.
\newblock \showarticletitle{The Moderating Influence of Privacy Concern on the
  Efficacy of Privacy Assurance Mechanisms for Building Trust: A
  Multiple-Context Investigation}. In {\em {ICIS} 2008 Proceedings}. Paris,
  France.
\newblock
\showURL{%
\url{http://aisel.aisnet.org/icis2008/7}}


\bibitem[\protect\citeauthoryear{Barocas and Nissenbaum}{Barocas and
  Nissenbaum}{2009}]%
        {Barocas2009}
{Solon Barocas} {and} {Helen Nissenbaum}. 2009.
\newblock \showarticletitle{On notice: The trouble with Notice and Consent}. In
  {\em Proceedings of the Engaging Data Forum: The First International Forum on
  the Application and Management of Personal Electronic Information}.
\newblock
\showURL{%
\url{http://www.nyu.edu/pages/projects/nissenbaum/papers/ED_SII_On_Notice.pdf}}


\bibitem[\protect\citeauthoryear{Becker and Pousttchi}{Becker and
  Pousttchi}{2012}]%
        {Becker2012}
{Laura Becker} {and} {Key Pousttchi}. 2012.
\newblock \showarticletitle{Social Networks: The Role of Users' Privacy
  Concerns}. In {\em Proceedings of the 14th International Conference on
  Information Integration and Web-based Applications \& Services} {\em (IIWAS
  '12)}. ACM, 187--195.
\newblock
\showISBNx{978-1-4503-1306-3}
\showDOI{%
\url{http://dx.doi.org/10.1145/2428736.2428767}}


\bibitem[\protect\citeauthoryear{Benisch, Kelley, Sadeh, and Cranor}{Benisch
  et~al\mbox{.}}{2011}]%
        {Benisch2011}
{Michael Benisch}, {Patrick~Gage Kelley}, {Norman Sadeh}, {and} {Lorrie~Faith
  Cranor}. 2011.
\newblock \showarticletitle{Capturing location-privacy preferences: quantifying
  accuracy and user-burden tradeoffs}.
\newblock {\em Personal Ubiquitous Computing\/} {15}, 7 (Oct. 2011), 679--694.
\newblock
\showISSN{1617-4909}
\showDOI{%
\url{http://dx.doi.org/10.1007/s00779-010-0346-0}}


\bibitem[\protect\citeauthoryear{Borcea-Pfitzmann, Pfitzmann, and
  Berg}{Borcea-Pfitzmann et~al\mbox{.}}{2011}]%
        {Borcea2011}
{Katrin Borcea-Pfitzmann}, {Andreas Pfitzmann}, {and} {Manuela Berg}. 2011.
\newblock \showarticletitle{Privacy 3.0 := Data Minimization + User Control +
  Contextual Integrity}.
\newblock {\em Information Technology\/} {53}, 1 (2011), 34--40.
\newblock
\showISSN{1611-2776}
\showDOI{%
\url{http://dx.doi.org/10.1524/itit.2011.0622}}


\bibitem[\protect\citeauthoryear{Brenner and Smith}{Brenner and Smith}{2013}]%
        {onlineArticle2013}
{Joanna Brenner} {and} {Aaron Smith}. 2013.
\newblock \showarticletitle{72\% of Online Adults are Social Networking Site
  Users}.
\newblock {\em PewResearch Internet Project\/} (2013).
\newblock


\bibitem[\protect\citeauthoryear{Compa{\~n}{\'o} and Lusoli}{Compa{\~n}{\'o}
  and Lusoli}{2010}]%
        {Compano2010}
{Ram{\'o}n Compa{\~n}{\'o}} {and} {Wainer Lusoli}. 2010.
\newblock \showarticletitle{The Policy Maker's Anguish: Regulating Personal
  Data Behavior Between Paradoxes and Dilemmas}.
\newblock In {\em Economics of Information Security and Privacy}, {Tyler
  Moore}, {David Pym}, {and} {Christos Ioannidis} (Eds.). Springer {US},
  169--185.
\newblock
\showISBNx{978-1-4419-6967-5}
\showURL{%
\url{10.1007/978-1-4419-6967-5_9}}


\bibitem[\protect\citeauthoryear{Consolvo, Smith, Matthews, LaMarca, Tabert,
  and Powledge}{Consolvo et~al\mbox{.}}{2005}]%
        {Consolvo2005}
{Sunny Consolvo}, {Ian~E. Smith}, {Tara Matthews}, {Anthony LaMarca}, {Jason
  Tabert}, {and} {Pauline Powledge}. 2005.
\newblock \showarticletitle{Location Disclosure to Social Relations: Why, when,
  \& What People Want to Share}. In {\em Proceedings of the SIGCHI Conference
  on Human Factors in Computing Systems} {\em (CHI '05)}. ACM, 81--90.
\newblock
\showISBNx{1-58113-998-5}


\bibitem[\protect\citeauthoryear{Culnan}{Culnan}{1993}]%
        {Culnan1993}
{Mary~J. Culnan}. 1993.
\newblock \showarticletitle{"How Did They Get My Name?": An Exploratory
  Investigation of Consumer Attitudes toward Secondary Information Use}.
\newblock {\em {MIS} Quarterly\/} {17}, 3 (1993), 341--363.
\newblock
\showISSN{0276-7783}
\showDOI{%
\url{http://dx.doi.org/10.2307/249775}}


\bibitem[\protect\citeauthoryear{DiMicco, Millen, Geyer, Dugan, Brownholtz, and
  Muller}{DiMicco et~al\mbox{.}}{2008}]%
        {DiMicco2008}
{Joan DiMicco}, {David~R. Millen}, {Werner Geyer}, {Casey Dugan}, {Beth
  Brownholtz}, {and} {Michael Muller}. 2008.
\newblock \showarticletitle{Motivations for Social Networking at Work}. In {\em
  Proceedings of the 2008 ACM Conference on Computer Supported Cooperative
  Work} {\em (CSCW '08)}. ACM, 711--720.
\newblock
\showISBNx{978-1-60558-007-4}
\showDOI{%
\url{http://dx.doi.org/10.1145/1460563.1460674}}


\bibitem[\protect\citeauthoryear{Dong, Jin, and Knijnenburg}{Dong
  et~al\mbox{.}}{2015}]%
        {DongJKicwsm}
{Cailing Dong}, {Hongxia Jin}, {and} {Bart~P. Knijnenburg}. 2015.
\newblock \showarticletitle{Predicting Privacy Behavior on Online Social
  Networks}. In {\em Proceedings of the Ninth International Conference on Web
  and Social Media, {ICWSM} 2015, University of Oxford, Oxford, UK, May 26-29,
  2015}. 91--100.
\newblock
\showURL{%
\url{http://www.aaai.org/ocs/index.php/ICWSM/ICWSM15/paper/view/10554}}


\bibitem[\protect\citeauthoryear{Ellison, Steinfield, and Lampe}{Ellison
  et~al\mbox{.}}{2007}]%
        {ellison2007}
{Nicole Ellison}, {Charles Steinfield}, {and} {Cliff Lampe}. 2007.
\newblock \showarticletitle{The Benefits of Facebook ``Friends: " Social
  Capital and College Students' Use of Online Social Network Sites}.
\newblock {\em Journal of Computer-Mediated Communication\/} (2007),
  1143--1168.
\newblock


\bibitem[\protect\citeauthoryear{Fang and LeFevre}{Fang and LeFevre}{2010}]%
        {Fang2010}
{Lujun Fang} {and} {Kristen LeFevre}. 2010.
\newblock \showarticletitle{Privacy Wizards for Social Networking Sites}. In
  {\em Proceedings of the 19th International Conference on World Wide Web} {\em
  (WWW '10)}. ACM, 351--360.
\newblock
\showISBNx{978-1-60558-799-8}


\bibitem[\protect\citeauthoryear{Gong, Xu, Huang, Mittal, Stefanov, Sekar, and
  Song}{Gong et~al\mbox{.}}{2011}]%
        {Gong2011}
{Neil~Zhenqiang Gong}, {Wenchang Xu}, {Ling Huang}, {Prateek Mittal}, {Emil
  Stefanov}, {Vyas Sekar}, {and} {Dawn Song}. 2011.
\newblock \showarticletitle{Evolution of Social-Attribute Networks:
  Measurements, Modeling, and Implications using Google+}. In {\em Proceedings
  of the 2012 ACM Conference on Internet Measurement Conference} {\em
  (IMC'11)}. 131--144.
\newblock


\bibitem[\protect\citeauthoryear{Gross and Acquisti}{Gross and
  Acquisti}{2005}]%
        {Gross2005}
{Ralph Gross} {and} {Alessandro Acquisti}. 2005.
\newblock \showarticletitle{Information Revelation and Privacy in Online Social
  Networks}. In {\em Proceedings of the 2005 ACM Workshop on Privacy in the
  Electronic Society} {\em (WPES '05)}. ACM, 71--80.
\newblock
\showISBNx{1-59593-228-3}
\showDOI{%
\url{http://dx.doi.org/10.1145/1102199.1102214}}


\bibitem[\protect\citeauthoryear{Hall, Frank, Holmes, Pfahringer, Reutemann,
  and Witten}{Hall et~al\mbox{.}}{2009}]%
        {weka2009}
{Mark Hall}, {Eibe Frank}, {Geoffrey Holmes}, {Bernhard Pfahringer}, {Peter
  Reutemann}, {and} {Ian~H. Witten}. 2009.
\newblock \showarticletitle{The WEKA Data Mining Software: An Update}.
\newblock {\em SIGKDD Explorations\/} {11}, 1 (2009).
\newblock


\bibitem[\protect\citeauthoryear{Harris, Westin, and Associates}{Harris
  et~al\mbox{.}}{1998}]%
        {Westin1998}
{L. Harris}, {A.F. Westin}, {and} {Associates}. 1998.
\newblock \showarticletitle{Personalized Marketing and Privacy on The Net: What
  Consumers Want}.
\newblock {\em Privacy and American Business Newsletter.\/} (1998).
\newblock


\bibitem[\protect\citeauthoryear{Ho, Maiga, and A{\"\i}meur}{Ho
  et~al\mbox{.}}{2009}]%
        {HoMA09}
{Ai Ho}, {Abdou Maiga}, {and} {Esma A{\"\i}meur}. 2009.
\newblock \showarticletitle{Privacy protection issues in social networking
  sites.}. In {\em AICCSA}. IEEE, 271--278.
\newblock
\showISBNx{978-1-4244-3807-5}


\bibitem[\protect\citeauthoryear{Hsu}{Hsu}{2006}]%
        {Hsu2006}
{Chiung-wen Hsu}. 2006.
\newblock \showarticletitle{Privacy concerns, privacy practices and web site
  categories: Toward a situational paradigm}.
\newblock {\em Online Information Review\/} {30}, 5 (2006), 569--586.
\newblock
\showISSN{1468-4527}
\showDOI{%
\url{http://dx.doi.org/10.1108/14684520610706433}}


\bibitem[\protect\citeauthoryear{Johnson, Egelman, and Bellovin}{Johnson
  et~al\mbox{.}}{2012}]%
        {Johnson2012}
{Maritza Johnson}, {Serge Egelman}, {and} {Steven~M. Bellovin}. 2012.
\newblock \showarticletitle{Facebook and privacy: it's complicated}. In {\em
  Proc. of the 8th Symposium on Usable Privacy and Security}. {ACM},
  Pittsburgh, {PA}.
\newblock
\showISBNx{978-1-4503-1532-6}
\showDOI{%
\url{http://dx.doi.org/10.1145/2335356.2335369}}


\bibitem[\protect\citeauthoryear{Joinson}{Joinson}{2008}]%
        {Joinson2008}
{Adam~N. Joinson}. 2008.
\newblock \showarticletitle{Looking at, Looking Up or Keeping Up with People?:
  Motives and Use of Facebook}. In {\em Proceedings of the SIGCHI Conference on
  Human Factors in Computing Systems} {\em (CHI '08)}. ACM, 1027--1036.
\newblock
\showISBNx{978-1-60558-011-1}
\showDOI{%
\url{http://dx.doi.org/10.1145/1357054.1357213}}


\bibitem[\protect\citeauthoryear{Kairam, Brzozowski, Huffaker, and Chi}{Kairam
  et~al\mbox{.}}{2012}]%
        {Kairam2012}
{Sanjay Kairam}, {Mike Brzozowski}, {David Huffaker}, {and} {Ed Chi}. 2012.
\newblock \showarticletitle{Talking in circles: Selective Sharing in Google+}.
  In {\em Proceedings of the {SIGCHI} Conference on Human Factors in Computing
  Systems}. {ACM} Press, Austin, {TX}, 1065--1074.
\newblock
\showISBNx{9781450310154}
\showDOI{%
\url{http://dx.doi.org/10.1145/2207676.2208552}}


\bibitem[\protect\citeauthoryear{Knijnenburg}{Knijnenburg}{2013}]%
        {KnijnenburgDecisions2013}
{Bart~P Knijnenburg}. 2013.
\newblock \showarticletitle{Simplifying Privacy Decisions: Towards Interactive
  and Adaptive Solutions}. In {\em Proceedings of the Recsys 2013 Workshop on
  Human Decision Making in Recommender Systems (Decisions@ {RecSys}'13)}. Hong
  Kong, China, 40--41.
\newblock


\bibitem[\protect\citeauthoryear{Knijnenburg}{Knijnenburg}{2015}]%
        {Knijnenburg2015}
{Bart~Piet Knijnenburg}. 2015.
\newblock {\em A user-tailored approach to privacy decision support}.
\newblock Ph.{D}. University of California, Irvine, United States --
  California.
\newblock
\showURL{%
\url{http://search.proquest.com/docview/1725139739/abstract}}


\bibitem[\protect\citeauthoryear{Knijnenburg and Jin}{Knijnenburg and
  Jin}{2013}]%
        {KnijnenburgSigHCI2013}
{Bart~P Knijnenburg} {and} {Hongxia Jin}. 2013.
\newblock \showarticletitle{The Persuasive Effect of Privacy Recommendations}.
  In {\em Twelfth Annual Workshop on {HCI} Research in {MIS}}. Milan, Italy.
\newblock
\showURL{%
\url{http://aisel.aisnet.org/sighci2013/16}}


\bibitem[\protect\citeauthoryear{Knijnenburg and Kobsa}{Knijnenburg and
  Kobsa}{2013}]%
        {Knijnenburg2013}
{Bart~P. Knijnenburg} {and} {Alfred Kobsa}. 2013.
\newblock \showarticletitle{Making Decisions About Privacy: Information
  Disclosure in Context-Aware Recommender Systems}.
\newblock {\em ACM Trans. Interact. Intell. Syst.\/} {3}, 3, Article 20 (Oct.
  2013), 23 pages.
\newblock
\showISSN{2160-6455}
\showDOI{%
\url{http://dx.doi.org/10.1145/2499670}}


\bibitem[\protect\citeauthoryear{Knijnenburg and Kobsa}{Knijnenburg and
  Kobsa}{2014}]%
        {Knijnenburg2014}
{B.~P. Knijnenburg} {and} {A. Kobsa}. 2014.
\newblock \showarticletitle{Increasing Sharing Tendency Without Reducing
  Satisfaction: Finding the Best Privacy-Settings User Interface for Social
  Networks}. In {\em {ICIS} 2014 Proceedings}. Auckland, New Zealand.
\newblock


\bibitem[\protect\citeauthoryear{Knijnenburg, Kobsa, and Jin}{Knijnenburg
  et~al\mbox{.}}{2013a}]%
        {KnijnenburgJin2013}
{Bart~P. Knijnenburg}, {Alfred Kobsa}, {and} {Hongxia Jin}. 2013a.
\newblock \showarticletitle{Dimensionality of Information Disclosure Behavior}.
\newblock {\em International Journal of Human-Computer Studies\/} {71}, 12
  (Dec. 2013), 1144--1162.
\newblock
\showISSN{1071-5819}


\bibitem[\protect\citeauthoryear{Knijnenburg, Kobsa, and Jin}{Knijnenburg
  et~al\mbox{.}}{2013b}]%
        {BartJinLocation2013}
{Bart~P. Knijnenburg}, {Alfred Kobsa}, {and} {Hongxia Jin}. 2013b.
\newblock \showarticletitle{Preference-based Location Sharing: Are More Privacy
  Options Really Better?}. In {\em Proceedings of the SIGCHI Conference on
  Human Factors in Computing Systems} {\em (CHI '13)}. ACM, 2667--2676.
\newblock
\showISBNx{978-1-4503-1899-0}


\bibitem[\protect\citeauthoryear{Kyumin~Lee and Caverlee}{Kyumin~Lee and
  Caverlee}{2011}]%
        {Lee2011}
{Brian David~Eoff Kyumin~Lee} {and} {James Caverlee}. 2011.
\newblock \showarticletitle{Seven Months with the Devils: A Long-Term Study of
  Content Polluters on Twitter}. In {\em International AAAI Conference on
  Weblogs and Social Media (ICWSM)}.
\newblock


\bibitem[\protect\citeauthoryear{Lampe, Ellison, and Steinfield}{Lampe
  et~al\mbox{.}}{2008}]%
        {Lampe2008}
{Cliff Lampe}, {Nicole~B. Ellison}, {and} {Charles Steinfield}. 2008.
\newblock \showarticletitle{Changes in Use and Perception of Facebook}. In {\em
  Proceedings of the 2008 ACM Conference on Computer Supported Cooperative
  Work} {\em (CSCW '08)}. ACM, 721--730.
\newblock
\showISBNx{978-1-60558-007-4}


\bibitem[\protect\citeauthoryear{Lampe, Gray, Fiore, and Ellison}{Lampe
  et~al\mbox{.}}{2014}]%
        {Lampe2014}
{Cliff Lampe}, {Rebecca Gray}, {Andrew~T. Fiore}, {and} {Nicole Ellison}. 2014.
\newblock \showarticletitle{Help is on the Way: Patterns of Responses to
  Resource Requests on Facebook}. In {\em Proceedings of the 17th ACM
  Conference on Computer Supported Cooperative Work \& Social Computing} {\em
  (CSCW '14)}. ACM, 3--15.
\newblock
\showISBNx{978-1-4503-2540-0}
\showDOI{%
\url{http://dx.doi.org/10.1145/2531602.2531720}}


\bibitem[\protect\citeauthoryear{Laufer and Wolfe}{Laufer and Wolfe}{1977}]%
        {Laufer1977}
{Robert~S Laufer} {and} {Maxine Wolfe}. 1977.
\newblock \showarticletitle{Privacy as a Concept and a Social Issue: A
  Multidimensional Developmental Theory}.
\newblock {\em Journal of Social Issues\/} {33}, 3 (1977), 22--42.
\newblock
\showDOI{%
\url{http://dx.doi.org/10.1111/j.1540-4560.1977.tb01880.x}}


\bibitem[\protect\citeauthoryear{Lederer, Hong, Dey, and Landay}{Lederer
  et~al\mbox{.}}{2004}]%
        {Lederer2004}
{Scott Lederer}, {Jason~I. Hong}, {Anind~K. Dey}, {and} {James~A. Landay}.
  2004.
\newblock \showarticletitle{Personal privacy through understanding and action:
  five pitfalls for designers}.
\newblock {\em Personal and Ubiquitous Computing\/} {8}, 6 (2004), 440--454.
\newblock
\showISSN{1617-4909}
\showDOI{%
\url{http://dx.doi.org/10.1007/s00779-004-0304-9}}


\bibitem[\protect\citeauthoryear{Lederer, Mankoff, and Dey}{Lederer
  et~al\mbox{.}}{2003}]%
        {Lederer2003}
{Scott Lederer}, {Jennifer Mankoff}, {and} {Anind~K. Dey}. 2003.
\newblock \showarticletitle{Who Wants to Know What when? Privacy Preference
  Determinants in Ubiquitous Computing}. In {\em CHI '03 Extended Abstracts on
  Human Factors in Computing Systems} {\em (CHI EA '03)}. ACM, 724--725.
\newblock
\showISBNx{1-58113-637-4}
\showDOI{%
\url{http://dx.doi.org/10.1145/765891.765952}}


\bibitem[\protect\citeauthoryear{Lewis, Kaufman, and Christakis}{Lewis
  et~al\mbox{.}}{2008}]%
        {LewisKevin2008}
{Kevin Lewis}, {Jason Kaufman}, {and} {Nicholas Christakis}. 2008.
\newblock \showarticletitle{The taste for privacy: An analysis of college
  student privacy settings in an online social network}.
\newblock {\em Journal of Computer-Mediated Communication\/} {14}, 1 (2008),
  79--100.
\newblock


\bibitem[\protect\citeauthoryear{Liu, Gummadi, Krishnamurthy, and Mislove.}{Liu
  et~al\mbox{.}}{2011}]%
        {YabingLiu2011}
{Yabing Liu}, {Krishna~P. Gummadi}, {Balachander Krishnamurthy}, {and} {Alan
  Mislove.} 2011.
\newblock \showarticletitle{Analyzing facebook privacy settings: user
  expectations vs. reality}. In {\em In Proceedings of the 2011 ACM SIGCOMM
  conference on Internet measurement conference}. ACM, 61--70.
\newblock


\bibitem[\protect\citeauthoryear{Madden}{Madden}{2012}]%
        {Madden2012}
{Mary Madden}. 2012.
\newblock {\em Privacy management on social media sites}.
\newblock {T}echnical {R}eport. Pew Internet \& American Life Project, Pew
  Research Center, Washington, DC.
\newblock
\showURL{%
\url{http://www.pewinternet.org/2012/02/24/privacy-management-on-social-media-sites/}}


\bibitem[\protect\citeauthoryear{Madejski, Johnson, and Bellovin}{Madejski
  et~al\mbox{.}}{2012}]%
        {Madejski2012}
{M. Madejski}, {M. Johnson}, {and} {S.M. Bellovin}. 2012.
\newblock \showarticletitle{A study of privacy settings errors in an online
  social network}. In {\em 2012 {IEEE} International Conference on Pervasive
  Computing and Communications Workshops ({PERCOM} Workshops)}. Lugano,
  Switzerland, 340--345.
\newblock
\showDOI{%
\url{http://dx.doi.org/10.1109/PerComW.2012.6197507}}


\bibitem[\protect\citeauthoryear{Nissenbaum}{Nissenbaum}{2009}]%
        {Nissenbaum2009}
{Helen Nissenbaum}. 2009.
\newblock {\em Privacy in context: Technology, policy, and the integrity of
  social life}.
\newblock Stanford University Press, Stanford, {CA}.
\newblock


\bibitem[\protect\citeauthoryear{Nissenbaum}{Nissenbaum}{2011}]%
        {Nissenbaum2011}
{Helen Nissenbaum}. 2011.
\newblock \showarticletitle{A Contextual Approach to Privacy Online}.
\newblock {\em Daedalus\/} {140}, 4 (2011), 32--48.
\newblock
\showISSN{0011-5266}
\showDOI{%
\url{http://dx.doi.org/10.1162/DAED_a_00113}}


\bibitem[\protect\citeauthoryear{Norberg, Horne, and Horne.}{Norberg
  et~al\mbox{.}}{2007}]%
        {Norberg2007}
{Patricia~A. Norberg}, {Daniel~R. Horne}, {and} {David~A. Horne.} 2007.
\newblock \showarticletitle{The privacy paradox: Personal information
  disclosure intentions versus behaviors}.
\newblock {\em Journal of Consumer Affairs\/} {41}, 1 (2007), 100--126.
\newblock


\bibitem[\protect\citeauthoryear{Olson, Grudin, and Horvitz}{Olson
  et~al\mbox{.}}{2005}]%
        {Olson2005}
{Judith~S. Olson}, {Jonathan Grudin}, {and} {Eric Horvitz}. 2005.
\newblock \showarticletitle{A Study of Preferences for Sharing and Privacy}. In
  {\em CHI '05 Extended Abstracts on Human Factors in Computing Systems} {\em
  (CHI EA '05)}. ACM, 1985--1988.
\newblock
\showISBNx{1-59593-002-7}


\bibitem[\protect\citeauthoryear{Page, Kobsa, and Knijnenburg}{Page
  et~al\mbox{.}}{2012}]%
        {Page2012}
{Xinru Page}, {Alfred Kobsa}, {and} {Bart~P. Knijnenburg}. 2012.
\newblock \showarticletitle{Don't Disturb My Circles! Boundary Preservation Is
  at the Center of Location-Sharing Concerns}. In {\em Proceedings of the Sixth
  International {AAAI} Conference on Weblogs and Social Media}. Dublin,
  Ireland, 266--273.
\newblock
\showURL{%
\url{http://www.aaai.org/ocs/index.php/ICWSM/ICWSM12/paper/view/4679}}


\bibitem[\protect\citeauthoryear{Pallapa, Das, Di~Francesco, and Aura}{Pallapa
  et~al\mbox{.}}{2014}]%
        {Pallapa2014}
{Gautham Pallapa}, {Sajal~K. Das}, {Mario Di~Francesco}, {and} {Tuomas Aura}.
  2014.
\newblock \showarticletitle{Adaptive and context-aware privacy preservation
  exploiting user interactions in smart environments}.
\newblock {\em Pervasive and Mobile Computing\/}  {12} (June 2014), 232--243.
\newblock
\showISSN{1574-1192}
\showDOI{%
\url{http://dx.doi.org/10.1016/j.pmcj.2013.12.004}}


\bibitem[\protect\citeauthoryear{Ravichandran, Benisch, Kelley, and
  Sadeh}{Ravichandran et~al\mbox{.}}{2009}]%
        {Ravichandran2009}
{Ramprasad Ravichandran}, {Michael Benisch}, {Patrick Kelley}, {and} {Norman
  Sadeh}. 2009.
\newblock \showarticletitle{Capturing Social Networking Privacy Preferences:}.
\newblock In {\em Privacy Enhancing Technologies}, {Ian Goldberg} {and}
  {Mikhail Atallah} (Eds.). Lecture Notes in Computer Science, Vol. 5672.
  Springer Berlin / Heidelberg, 1--18.
\newblock
\showISBNx{978-3-642-03167-0}
\showURL{%
\url{10.1007/978-3-642-03168-7_1}}


\bibitem[\protect\citeauthoryear{Reports}{Reports}{2012}]%
        {CR2012}
{Consumer Reports}. 2012.
\newblock Facebook \& your privacy: Who sees the data you share on the biggest
  social network?
\newblock   (2012).
\newblock
\showURL{%
\url{http://www.consumerreports.org/cro/magazine/2012/06/facebook-your-privacy}}


\bibitem[\protect\citeauthoryear{Sadeh, Hong, Cranor, Fette, Kelley, Prabaker,
  and Rao}{Sadeh et~al\mbox{.}}{2009}]%
        {Sadeh2009}
{Norman Sadeh}, {Jason Hong}, {Lorrie Cranor}, {Ian Fette}, {Patrick Kelley},
  {Madhu Prabaker}, {and} {Jinghai Rao}. 2009.
\newblock \showarticletitle{Understanding and capturing people's privacy
  policies in a mobile social networking application}.
\newblock {\em Personal and Ubiquitous Computing\/} {13}, 6 (2009), 401--412.
\newblock
\showISSN{1617-4909}
\showDOI{%
\url{http://dx.doi.org/10.1007/s00779-008-0214-3}}


\bibitem[\protect\citeauthoryear{Sheehan}{Sheehan}{2002}]%
        {Sheehan2002}
{Kim~Bartel Sheehan}. 2002.
\newblock \showarticletitle{Toward a typology of Internet users and online
  privacy concerns}.
\newblock {\em The Information Society\/} {18}, 1 (2002).
\newblock


\bibitem[\protect\citeauthoryear{Smith, Goldstein, and Johnson}{Smith
  et~al\mbox{.}}{2013}]%
        {Smith2013}
{N.~Craig Smith}, {Daniel~G. Goldstein}, {and} {Eric~J. Johnson}. 2013.
\newblock \showarticletitle{Choice Without Awareness: Ethical and Policy
  Implications of Defaults}.
\newblock {\em Journal of Public Policy \& Marketing\/} {32}, 2 (2013),
  159--172.
\newblock
\showISSN{07439156}
\showDOI{%
\url{http://dx.doi.org/10.1509/jppm.10.114}}


\bibitem[\protect\citeauthoryear{Solove}{Solove}{2013}]%
        {Solove2013}
{Daniel~J. Solove}. 2013.
\newblock \showarticletitle{Privacy Self-Management and the Consent Dilemma}.
\newblock {\em Harvard Law Review\/}  {126} (2013), 1880--1903.
\newblock
\showURL{%
\url{http://papers.ssrn.com/abstract=2171018}}


\bibitem[\protect\citeauthoryear{Swets}{Swets}{1988}]%
        {Swets1988}
{J.~A. Swets}. 1988.
\newblock \showarticletitle{Measuring the accuracy of diagnostic systems}.
\newblock {\em Science (New York, N.Y.)\/} {240}, 4857 (June 1988), 1285--1293.
\newblock
\showISSN{0036-8075}


\bibitem[\protect\citeauthoryear{Tang, Hong, and Siewiorek}{Tang
  et~al\mbox{.}}{2012}]%
        {Tang2012}
{Karen Tang}, {Jason Hong}, {and} {Dan Siewiorek}. 2012.
\newblock \showarticletitle{The Implications of Offering More Disclosure
  Choices for Social Location Sharing}. In {\em Proceedings of the SIGCHI
  Conference on Human Factors in Computing Systems} {\em (CHI '12)}. ACM,
  391--394.
\newblock
\showISBNx{978-1-4503-1015-4}


\bibitem[\protect\citeauthoryear{Tang, Lin, Hong, Siewiorek, and Sadeh}{Tang
  et~al\mbox{.}}{2010}]%
        {Tang2010}
{Karen Tang}, {Jialiu Lin}, {Jason Hong}, {Daniel Siewiorek}, {and} {Norman
  Sadeh}. 2010.
\newblock \showarticletitle{Rethinking location sharing: exploring the
  implications of social-driven vs. purpose-driven location sharing}. In {\em
  Proc. {UbiComp} 2010}. Copenhagen, Denmark, 85--04.
\newblock
\showDOI{%
\url{http://dx.doi.org/10.1145/1864349.1864363}}


\bibitem[\protect\citeauthoryear{Taylor}{Taylor}{2003}]%
        {Taylor2003}
{Humphrey Taylor}. 2003.
\newblock \showarticletitle{Most people are ``privacy pragmatists'' who, while
  concerned about privacy, will sometimes trade it off for other benefits}.
\newblock {\em The Harris Poll\/} {17}, 19 (2003).
\newblock


\bibitem[\protect\citeauthoryear{Toch, Cranshaw, Drielsma, Tsai, Kelley,
  Springfield, Cranor, Hong, and Sadeh}{Toch et~al\mbox{.}}{2010}]%
        {Toch2010}
{Eran Toch}, {Justin Cranshaw}, {Paul~Hankes Drielsma}, {Janice~Y. Tsai},
  {Patrick~Gage Kelley}, {James Springfield}, {Lorrie Cranor}, {Jason Hong},
  {and} {Norman Sadeh}. 2010.
\newblock \showarticletitle{Empirical models of privacy in location sharing}.
  In {\em Proc. of the 12th {ACM} intl. conference on Ubiquitous computing}.
  {ACM} Press, Copenhagen, Denmark, 129--138.
\newblock
\showISBNx{978-1-60558-843-8}
\showDOI{%
\url{http://dx.doi.org/10.1145/1864349.1864364}}


\bibitem[\protect\citeauthoryear{Watson, Besmer, and Lipford}{Watson
  et~al\mbox{.}}{2012}]%
        {Watson2012}
{Jason Watson}, {Andrew Besmer}, {and} {Heather~Richter Lipford}. 2012.
\newblock \showarticletitle{+Your circles: sharing behavior on Google+}. In
  {\em Proc. of the 8th Symposium on Usable Privacy and Security}. {ACM},
  Pittsburgh, {PA}.
\newblock
\showISBNx{978-1-4503-1532-6}
\showDOI{%
\url{http://dx.doi.org/10.1145/2335356.2335373}}


\bibitem[\protect\citeauthoryear{Woodruff, Pihur, Consolvo, Schmidt,
  Brandimarte, and Acquisti}{Woodruff et~al\mbox{.}}{2014}]%
        {Woodruff2014}
{Allison Woodruff}, {Vasyl Pihur}, {Sunny Consolvo}, {Lauren Schmidt}, {Laura
  Brandimarte}, {and} {Alessandro Acquisti}. 2014.
\newblock \showarticletitle{Would a privacy fundamentalist sell their {DNA} for
  \$1000... if nothing bad happened as a result? {The} {Westin} categories,
  behavioral intentions, and consequences}. In {\em Symposium on {Usable}
  {Privacy} and {Security} ({SOUPS})}.
\newblock
\showURL{%
\url{https://www.usenix.org/system/files/conference/soups2014/soups14-paper-woodruff.pdf}}


\bibitem[\protect\citeauthoryear{Xu, Dinev, Smith, and Hart}{Xu
  et~al\mbox{.}}{2008}]%
        {Xu2008}
{H. Xu}, {T. Dinev}, {H.~J Smith}, {and} {P. Hart}. 2008.
\newblock \showarticletitle{Examining the formation of individual's privacy
  concerns: Toward an integrative view}. In {\em {ICIS} 2008 Proceedings}.
  Paris, France.
\newblock


\bibitem[\protect\citeauthoryear{Zickuhr}{Zickuhr}{2012}]%
        {Zickuhr2012}
{Kathryn Zickuhr}. 2012.
\newblock {\em Three-quarters of smartphone owners use location-based
  services}.
\newblock {T}echnical {R}eport. Pew Research Center.
\newblock
\showURL{%
\url{http://pewinternet.org/~/media//Files/Reports/2012/PIP_Location_based_services_2012_Report.pdf}}


\end{thebibliography}
                             % Sample .bib file with references that match those in
                             % the 'Specifications Document (V1.5)' as well containing
                             % 'legacy' bibs and bibs with 'alternate codings'.
                             % Gerry Murray - March 2012

% History dates
%\received{February 2007}{March 2009}{June 2009}
%
%% Electronic Appendix
%\elecappendix
%
%\medskip
%
%\section{This is an example of Appendix section head}
%
%Channel-switching time is measured as the time length it takes for
%motes to successfully switch from one channel to another. This
%parameter impacts the maximum network throughput, because motes
%cannot receive or send any packet during this period of time, and it
%also affects the efficiency of toggle snooping in MMSN, where motes
%need to sense through channels rapidly.
%
%By repeating experiments 100 times, we get the average
%channel-switching time of Micaz motes: 24.3 $\mu$s. We then conduct
%the same experiments with different Micaz motes, as well as
%experiments with the transmitter switching from Channel 11 to other
%channels. In both scenarios, the channel-switching time does not have
%obvious changes. (In our experiments, all values are in the range of
%23.6 $\mu$s to 24.9 $\mu$s.)
%
%\section{Appendix section head}
%
%The primary consumer of energy in WSNs is idle listening. The key to
%reduce idle listening is executing low duty-cycle on nodes. Two
%primary approaches are considered in controlling duty-cycles in the
%MAC layer.

\end{document}